\documentclass[twocolumn,aps, prl,showpacs,preprintnumbers,amsmath,amssymb]{revtex4}
\usepackage{graphicx}
\usepackage{dcolumn}
\usepackage{bm}
\usepackage{sidecap}

\def\ket#1{\lvert#1\rangle}
\def\bra#1{\langle #1\lvert}

\begin{document}


\title{Engineering of triply entangled states in a single-neutron system }
\author{Yuji Hasegawa$^{1,2}$}
\author{Rudolf Loidl$^{1,3}$}
\author{Gerald Badurek$^1$}
\author{Stephan Sponar$^1$}
\author{Helmut Rauch$^{1,3}$}
\affiliation{%
$^1$Atominstitut der \"{O}sterreichischen Universit\"{a}ten, Stadionallee 2, A-1020 Wien, Austria\\
$^2$PRESTO, Japan Science and Technology Agency (JST), Kawaguchi,
Saitama, Japan \\
$^3$Institut Laue-Langevin, B.P. 156, F-38042 Grenoble Cedex
9, France  }

\date{  submitted to PRL}
\begin{abstract}

We implemented a triply entangled
Greenberger-Horne-Zeilinger(GHZ)-like state and coherently
manipulated the spin, path, and energy degrees of freedom in a
single-neutron system. The GHZ-like state was analyzed with an
inequality derived by Mermin: we determined the four expectation values and finally obtained $M=2.558\pm0.004\nleq2$, which exhibits a clear violation of the noncontextual assumption
and confirms quantum contextuality.

\end{abstract}


\maketitle
Einstein, Podolsky, and Rosen argued that quantum mechanics (QM) is
not a complete theory in the sense that some results which can be
predicted from remote measurements are not described by the theory
[1]. Bell showed that local hidden variable theories (LHVTs) satisfy
some inequalities which are violated by QM [2], thus QM cannot be
completed with LHVTs. Experimental violations of Bell inequalities on
bipartite systems have been observed with two-photons
\cite{101}, two-ions \cite{102}, atom-photon systems \cite{103},
and two-hadrons \cite{104}. Moreover, Bell-like
inequalities can be tested using different degrees of freedom of
single-particle systems. In this scenario, the violation of the inequality
does not prove the impossibility of LHVTs, but the impossibility of
noncontextual hidden variable theories (NCHVTs) \cite{12,11}.
In NCHVTs the result of a measurement $\hat{A}$ is predetermined and
is not affected by other previous (or simultaneous) measurements,
carried out on the same individual system, of
any observables  mutually commuting with $\hat{A}$. Experimental violations
of Bell-like inequalities with two degrees of freedom of single
neutrons have been observed \cite{10}.

Even more apparent conflicts between predictions by QM
and LHVTs are found by Greenberger, Horne, and Zeilinger:
entangled states of three or more separated systems can lead to
predictions nonstatistically in contradiction to each other \cite{13}.
Indeed, Mermin showed that this conflict can be converted into a larger
violation of a Bell-like inequality between three or more separated systems \cite{27}.
Experimental test of these inequalities were reported, e.g., with the use of
three and four photons \cite{15,16} and four ions \cite{105}: among these, tests of quantum non-locality on many-particle generalizations of the GHZ triplet are particularly appealing \cite{21}. A natural question is whether a violation of
Mermin-like inequalities can be observed also on single-particle systems. The
interest of this violation goes beyond the technical challenge
of preparing GHZ-like entangled states using three degrees of
freedom of a single-particle system and the capability of measuring the
corresponding observables. The violation of the Mermin-like
inequality is interesting in itself since it is more robust to noise
than previous violations of bipartite Bell-like inequalities and
thus emphasizes the conflict between QM and NCHVTs.

Starting from a demonstration of a violation of Bell-like inequality \cite{10}, several neutron optical experiments were accomplished using Bell-like states, with entanglement of two, i.e., the spin and the path, degrees of freedom of neutrons \cite{23,24}. Recently we developed a coherent-manipulation method of a neutron's energy, i.e., total energy of neutrons given by the sum of kinematic and potential energies \cite{25}. This technique accompanied by phase manipulations \cite{26} allows us to add one more degree of freedom to be entangled: a triply entangled GHZ-like state in a single neutron system is generated and manipulated.
Here we report the first preparation of a GHZ-like state using three degrees of freedom of a single neutron, i.e., two internal degrees of freedom (the spin
and the energy) and one external one (the path taken by the neutron in
an interferometer setup), and the first violation of a Mermin-like
inequality with a single-particle system. General descriptions of perfect crystal neutron interferometer experiments can be found in a book \cite{28}.

\begin{figure}
  \centering
  \includegraphics [width=85mm]{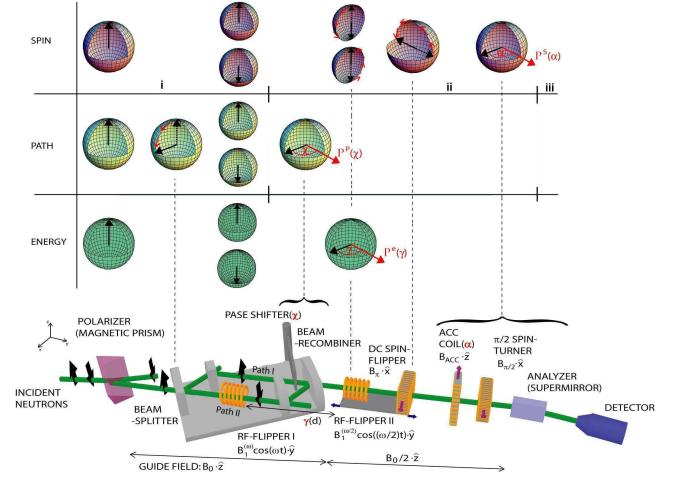}
  \caption{Schematic view of the main experimental setup (not to scale) for the preparation and analysis of triply-entangled states in a single-neutron system together with Bloch-sphere descriptions to depict evolutions of each quantum state, i.e., the spin, path and energy degrees of freedom. The experiment consist of three stages. (i) Preparation of a triply entangled GHZ-like state $\ket{\Psi^{GHZ}_N}$: the state of neutron is the 50/50 superposition of $\ket{\uparrow} \otimes \ket{\textrm{I}} \otimes \ket{E_0}$ (all states on the north poles) and $\ket{\downarrow} \otimes \ket{\textrm{II}} \otimes \ket{E_0-\hbar\omega}$ (all states on the south poles of the Bloch-spheres). (ii)Manipulation of the relative phases followed by projection measurements: the directions of the projection measurements $P^j$ are depicted by thick red arrows in Bloch-spheres. (iii) Detection: numbers of neutrons $N(\chi ; \alpha ; \gamma)$ are counted.
  }
  \label{Setup}
\end{figure}

In the neutron interferometer experiments accompanied by two radio-frequency (RF) oscillating fields (see Fig.\ref{Setup}), the total state consists of the neutron state $\lvert \Psi_N \rangle$ and the two RF fields, $\ket{\alpha_\omega}$ and $\ket{\alpha_{\omega /2}}$ represented by coherent states: $\ket{\Psi_{tot}} = \ket{\alpha_\omega  \rangle \otimes |\alpha_{\omega /2} \rangle \otimes |\Psi_N}$ \cite{25}. At first, the incident neutron was polarized, e.g., to up, denoted by $\ket{\uparrow}$: all states, corresponding to the spin, path and energy degrees of freedom, are represented by the north-pole points of the Bloch-sphere in Fig.1. In passing through the first plate (the beam splitter) of the interferometer, the state describing neutron's path is transformed into a 50/50 superposition of path-I ($\ket{\textrm{I}}$) and path-II ($\ket{\textrm{II}}$) states. Therefore, the corresponding state lies on the equator of the path Bloch-sphere.  In the interferometer, a RF spin-flipper operated with frequency $\omega$ was inserted in the path II, where the spin-flip process by a time-dependent interaction induces energy transitions from the initial energy state $\ket{E_0}$ to states $\ket{E_0\!-\!\hbar\omega}$ by photon exchange: up-spin $\ket{\uparrow}$ only in path-II was flipped to down-spin $\ket{\downarrow}$, thus losing energy by $\hbar\omega$ \cite{29}. Consequently, one can generate the state of neutrons in a triply entangled GHZ-like state, given by
\begin{equation}
 \label{eq:eq1}
\ket{\Psi^{GHZ}_N}=\frac{1}{\sqrt{2}}\{\ket{\uparrow} \otimes \ket{\textrm{I}} \otimes \ket{E_0} + \ket{\downarrow} \otimes \ket{\textrm{II}} \otimes \ket{E_0-\hbar\omega} \}.
\end{equation}
Note that, in this GHZ-like state, a product state $\ket{\uparrow} \otimes \ket{\textrm{I}} \otimes \ket{E_0}$, all states on the north poles, and another product state $\ket{\downarrow} \otimes \ket{\textrm{II}} \otimes \ket{E_0-\hbar\omega}$, all states on the south poles of the Bloch-spheres, are superposed.
Here, the state of neutrons is characterized by three,
i.e., the spin, path and energy, degrees of freedom,
which are simply described by two-level quantum systems such as
\begin{equation}
 \label{eq:eq2}
\left\{{\begin{array}{lll}
          \ket{\Psi_{spin}}=\{ \ket{\uparrow}, \ket{\downarrow} \} \\
          \ket{\Psi_{path}}=\{ \ket{\textrm{I}}, \ket{\textrm{II}} \} \\
          \ket{\Psi_{energy}}=\{ \ket{E_0-\hbar\omega}, \ket{E_0} \}.
        \end{array}
} \right.
\end{equation}
In this simple description, all subspaces are effectively spanned by orthogonal two-bases. It is noting here that the energy subspace (and the momentum subspace for the path degree of freedom) is(are), in principle, not a discrete two-level but has(have) a continuous structure. Observable of one subspace commute with that of a different subspace, which justifies the derivation of a Mermin-like inequality due to NCHVTs.

The important operations of each degree of freedom in the experiments were phase manipulations between each of two bases. (a) The spin-phase $\alpha$ was adjusted by a magnetic field oriented along the quantisation axis, i.e., +z direction, tuned by an 'accelerator' DC coil. In reality, the change of the Larmor frequency $\Delta \omega_L$ results in a phase shift $\alpha=\Delta \omega_L T_1$, where $T_1$ is the propagation time through the 'accelerator' coil. (b) The phase manipulation of the path subspace was accomplished with the use of an auxiliary phase shifter made of a parallel-sided Si plate 5mm in thickness. In this case, the phase shift $\chi$ was given by $\chi=N b_c \lambda D$, with the atom density $N$, the coherent scattering length $b_c$, the wavelength of the beam $\lambda$, and the thickness of the plate $D$. (b) There is a suitable method for a phase manipulation of the energy degree of freedom, which is known as a zero-field precession \cite{30}: when two RF flippers (operated at a frequency  $\omega_r$) are set in serial, the former induces the energy difference $\pm \hbar\omega_r$ until the latter, resulting in the phase difference $\gamma=2\omega_r T_2$, where $T_2$ is the propagation time between the flippers. In particular, an experimentally convenient method to manipulate individually the Larmor phase   and the zero-field phase $\gamma$ was found and reported in \cite{26}.

Since perfect correlations (or anti-correlations) can not be observed in real experiments, one should use an inequality in order to clarify peculiarities of the triply entangled GHZ-like state. Mermin analyzed the GHZ argument in detail and derived an inequality suitable for experimental tests to distinguish between predictions by QM and by LHVTs \cite{27}. Assuming a tripartite system and taking the assumption in the conditionally independent form (represented by the Eq.(5) in \cite{27}) due to NCHVTs instead of LHVTs, one can obtain the border for a sum of expectation values of certain product observables, which is to be tested in the experiment. The sum of expectation values $M$ is defined as
\begin{equation}
 \label{eq:eq3}
M\!=\!E[\sigma^p_x \sigma^s_x  \sigma^e_x]
-\!E[\sigma^p_x  \sigma^s_y  \sigma^e_y]
-\!E[\sigma^p_y  \sigma^s_x  \sigma^e_y]
-\!E[\sigma^p_y  \sigma^s_y  \sigma^e_x]
\end{equation}
where $\!E[\ldots]$, $\sigma^p_j$, $\sigma^s_j$, and $\sigma^e_j$ represent expectation values, and  Pauli operators for the two-level systems in the path, spin, and energy subspaces, respectively. NCHVTs set a strict limit for the maximum possible value of 2. In contrast, quantum theory predicts an upper bound of 4: any measured value of $M$ that is larger than 2 decides in favor of quantum contextuality. A violation of up to factor of 2 is expected with a triply entangled GHZ-like state $\ket{\Psi^{GHZ}_N}$.

\begin{SCfigure*}
  \centering
  \includegraphics [width=135mm]{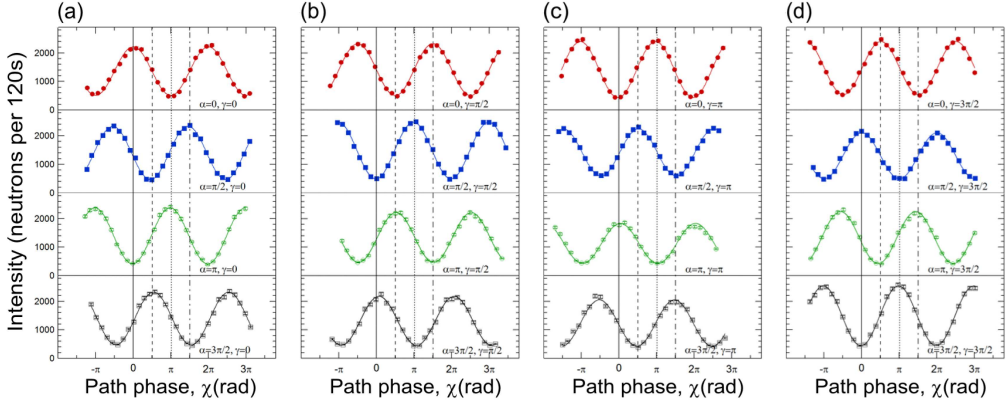}
  \caption{Typical intensity modulations obtained by varying the path-phase $\chi$. The phases $\alpha$ and $\gamma$, for the spin and the energy, respectively, are tuned at $0$, $\pi /2$, $\pi$ and $3\pi /2$ in order to accomplish projection measurements of $\hat P (0)$, $\hat P (\pi /2)$, $\hat P (\pi)$, and $\hat P (3\pi /2)$ for both: (a) $\gamma=0$, (b) $\gamma=\pi /2$ (c) $\gamma=\pi$ (d) $\gamma=3\pi /2$ and in each $\gamma$ setting $\alpha$ was set at $0$, $\pi /2$, $\pi$ and $3\pi /2$ (upper to lower panels).
    }
\label{Osci}
\end{SCfigure*}

In order to test the Mermin-like inequality, one should determine four expectation values for joint measurements of three, i.e., spin, path and energy, observables. We decided to extend the strategy used in the measurement of the Peres-Mermin proof of the Kochen-Specker theorem \cite{200}: successive measurements of three degrees of freedom were carried out.
 Projection operators $\hat P^l (\phi)$ to the state $\frac{1}{\sqrt{2}}\{ \ket{\Psi_l}+e^{i\phi}\ket{\Psi^{\perp}_l} \}$ (l=path, spin, energy) were involved in the experiments: $\sigma^l_x=\hat P^l (0)-\hat P^l (\pi)$ and $\sigma^l_y=\hat P^l (\pi/2)-\hat P^l (3\pi /2)$. Note that these projection operators differ only in phase between the two orthogonal states, which was experimentally very convenient since one only needs phase manipulations: directions of all projection measurements are depicted by thick red arrows in Bloch-spheres in Fig.1 (they all lie on the equatorial plane).
 In practice, each expectation value was determined by a combination of eight count rates in a single detector with appropriate phase settings, for instance,
\begin{equation}
 \label{eq:eq4}
 {\begin{array}{lll}
\hspace{4mm} E[\sigma^p_x  \sigma^s_y  \sigma^e_y] \\
\!= \!\bra{\Psi}[\hat P^p (0) \!-\!\!\hat P^p (\pi)] [\hat P^s (\frac{\pi}{2})\!-\!\!\hat P^s (\frac{3\pi}{2})][\hat P^e (\frac{\pi}{2})\!-\!\!\hat P^e (\frac{3\pi}{2})]\ket{\Psi} \\
\!=\!\frac{N(0:\frac{\pi}{2}:\frac{\pi}{2})-N(\pi:\frac{\pi}{2}:\frac{\pi}{2})-N(0:\frac{3\pi}{2}:\frac{\pi}{2})...+N(0:\frac{3\pi}{2}:\frac{3\pi}{2})-N(\pi:\frac{3\pi}{2}:\frac{3\pi}{2})}
     {{N(0:\frac{\pi}{2}:\frac{\pi}{2})+N(\pi:\frac{\pi}{2}:\frac{\pi}{2})+N(0:\frac{3\pi}{2}:\frac{\pi}{2})...+N(0:\frac{3\pi}{2}:\frac{3\pi}{2})+N(\pi:\frac{3\pi}{2}:\frac{3\pi}{2})}}
\end{array}}
\end{equation}
where $N(\chi ; \alpha ; \gamma)=\langle \Psi \lvert {\hat P^p (\chi)} \cdot {\hat P^s (\alpha)} \cdot {\hat P^e (\gamma)} \lvert \Psi \rangle$ denotes the count rate with the path phase $\chi$, the spin phase $\alpha$, and the energy phase $\gamma$.

	The experiment was carried out at the perfect-crystal neutron-interferometer beam line S18 at the high flux reactor at the Institute Laue Langevin (ILL). A schematic view of the main components of the experimental setup together with a Bloch-sphere description depicting evolutions of each degree of freedom is shown in Fig.\ref{Setup}. A silicon perfect-crystal monochromator was placed in the neutron guide to monochromatize the incident neutron beam to a mean wave length of  $\lambda_0$=1.92{\AA} with the monochromaticity ${\Delta\lambda}/{\lambda_0}\approx0.01$. The cross section of the incident beam was confined to $5\times5mm^2$. Magnetic prisms were used to polarize the incident beam vertically, before the beam enters a triple-Laue (LLL) interferometer. The interferometer was adjusted to give the 220 reflections. A parallel-sided Si plate was used as a phase shifter to tune the phase $\chi$ for the path degree of freedom. This phase shifter accompanied by the beam recombination in the interferometer enabled to realize a projection measurement with the operator $\hat P^p (\chi)$.

A fairly uniform magnetic guide field, $B_0$ in $+\hat z$ ($\sim$20G), was applied around the interferometer by a pair of water-cooled Helmholtz coils (not shown in Fig.\ref{Setup}). The first RF spin-flipper was located in this region, and its operational frequency was tuned to 58kHz. The GHZ-like state of neutrons $\ket{\Psi^{GHZ}_N}$ was generated by turning on this RF spin-flipper. Along the flight path after the interferometer, another fairly uniform magnetic guide field, $B'_0$ (at half strength $\sim$10G), was applied with another pair of water-cooled coils in Helmholtz geometry (also not depicted in Fig.\ref{Setup}). The second RF spin-flipper, tuned to the operational frequency of 29kHz, was placed in this region. This RF spin-flipper was mounted on a common translator together with a DC spin-flipper. The translation of the common basis allows one to tune the phase $\gamma$ of the energy degree of freedom independently \cite{26}; for instance, $\gamma=0$, $\pi/2$, $\pi$, $3\pi/2$ resulted in implementing  $\hat P^e (0)$, $\hat P^e (\pi/2)$, $\hat P^e (\pi)$, and $\hat P^e (3\pi/2)$. Note that the second RF spin-flipper, in practice, worked as an energy "recombiner" described as $\hat O^{(\textrm{E})}=\frac{1}{\sqrt{2}}\ket{E_0-\hbar\omega/2}\{\bra{E_0}+\bra{E_0-\hbar\omega}\}$. A spin-analyzer in the $+\hat z$ direction (a bent magnetically saturated bent Co-Ti supermirror) together with a $\pi/2$ spin-turner enabled the selection of neutrons in xy-plane (normal to the quantization axis). An accelerator coil, oriented in $B_{acc}$$+\hat z$ was used to adjust the spin phase $\alpha=0$, $\pi/2$, $\pi$, $3\pi/2$ accomplishing projection measurements of $\hat P^s (0)$, $\hat P^s (\pi/2)$, $\hat P^s (\pi)$, and $\hat P^s (3\pi/2)$.

By tuning the spin phase $\alpha$, and the energy phase $\gamma$ each at $0$, $\pi/2$, $\pi$ and $3\pi/2$, 16 independent path phase $\chi$ scan, i.e. oscillation measurements, was carried out for the determination of $M$. Typical oscillations are depicted in Fig.\ref{Osci}: intensities at indicated lines ($\chi=0$, $\pi/2$, $\pi$, $3\pi/2$) were used to determine the related expectation values. The contrasts of the oscillations were just below $70\%$, which was about the same as those with the empty interferometer and showed that all parameters could be manipulated effectively.

Measured intensity oscillations were fitted to sinusoidal curves by the least squares method and the four related expectation values were extracted. Statistical errors were estimated to $\pm0.001$ taking all fit-errors from single measurement curves into account. One set of measurements, consists of 32 oscillation measurements. (We recorded intensities with and without spin-flips at each phase shifter $\chi$ position, which allowed estimation and correction, if necessary, of the path-phase $\chi$ instability afterwards.) We measured 4 sets of such 32-oscillations to reduce statistical errors. During the analysis, we noticed that the statistical errors here are much smaller than those obtained in the Bell-like inequality experiments \cite{10}. This is due to the fact that the points used to determine expectation values are in the vicinity of the flat maxima or minima, e.g. $N(\chi:\alpha=0,\pi:\gamma=0,\pi)$ around $\chi=0, \pi$ on the solid and dotted lines in Fig.\ref{Osci}, which reflects robustness of the Mermin-like inequality and led to rather small statistical errors.
Four measurements were summed up as weighted averages and the final value and the error were determined. So, the final errors are the sum of systematic and statistical errors. (Systematic errors were mainly due to the path-phase $\chi$ instability, i.e., unwanted drifts of the $\chi$-phase, during the measurement.) We obtained four expectation values listed in Tab.1 together with settings of variables and the final $M$-value. In evaluating the Mermin-like inequality, $M$ was calculated to be $M=2.558\pm0.004$. This exhibits a clear violation $M\nleq2$ of the noncontextual border. The reduction from the ideal value of 4 is solely due to reduced contrast of the interference term from the interferometer, i.e. just below $70\%$.

\begin{table}
\caption{\label{tab:table1} Experimentally determined expectation values and the resulting M value.  }
\begin{ruledtabular}
\begin{tabular}{ccccr}
Observables&&Variables&&Values$\hspace{4mm}$\\
&$\chi$&$\alpha$&$\gamma$&\\
\hline
$\sigma^p_x\sigma^s_x\sigma^e_x$&0,$\pi$&0,$\pi$&0,$\pi$&$0.659(2)$\\
$\sigma^p_x\sigma^s_y\sigma^e_y$&0,$\pi$&$\pi/2$,$3\pi/2$&$\pi/2$,$3\pi/2$&$-0.603(2)$\\
$\sigma^p_y\sigma^s_x\sigma^e_y$&$\pi/2$,$3\pi/2$&0,$\pi$&$\pi/2$,$3\pi/2$&$-0.632(2)$\\
$\sigma^p_y\sigma^s_y\sigma^e_x$&$\pi/2$,$3\pi/2$&$\pi/2$,$3\pi/2$&0,$\pi$&$-0.664(2)$\\
\hline
&\hspace{8mm}$M$&\hspace{-8mm}$=$\hspace{2mm}$2.558$&\hspace{-13mm}$\pm$\hspace{2mm}$0.004$\\
\end{tabular}
\end{ruledtabular}
\end{table}

Our results with neutrons were obtained with detectors of more than 99\% efficiency. This experiment alone will not close all loopholes, e.g., a light-cone loophole is still remaining, but such a high efficiency of detectors for neutrons will help to consider physics of contextuality thoroughly. The use of entanglement of the energy degree of freedom is not limited to neutrons but easily applicable to other quantum systems. Furthermore, a coherent manipulation of energy degree of freedom can be extended to create artificial multi-level quantum system, e.g., in the order of $10^3$, in a single-particle system by applying a multiple-frequency energy-manipulation scheme in serial. Such a system could be used for quantum information processing.

In summary we have accomplished the demonstration of the violation of the Mermin-like inequality with the use of three, i.e., the spin, path and energy, degrees of freedom in a single-neutron system. The concept of entanglement is not limited between spatially-separated systems but also generally applicable between degrees of freedom. Here, as the first realization of triple entanglement in a single-particle system, the GHZ-like state was generated and analyzed. Now we are ready to proceed to investigate other triply entangled states, for instance, the W-state \cite{31} which is expected to be generated rather easily with a double-loop neutron interferometer \cite{32}.

We thank A. Cabello (Sevilla) and A. Hosoya (Tokyo) for his critical reading of the manuscript and appreciate discussions with E. Balcar, K. Durstberger-Rennhofer, J. Klepp (Vienna) and S. Filipp (Zurich). This work has been supported partly by the Japanese Science and Technology Agency, the Austrian Fonds zur F\"{o}derung der Wissenschaftlichen Forschung (No. P21193-N20).


\begin{thebibliography}{99}
\bibitem{1} A. Einstein, A. Podolsky, and N. Rosen, Phys. Rev.  \textbf{47}, 777 (1935).
\bibitem{2} J.S. Bell, Physics  \textbf{1}, 195 (1964); J.F. Clauser, and A. Shimony, Rep. Prog. Phys.  \textbf{41}, 1881 (1978); R.A. Bertlmann and A. Zeilinger (Eds.), \textit{Quantum [Un]speakables} (Springer Verlag, Berlin-Heidelberg, 2002).
\bibitem{101} for instancce, A. Aspect, J. Dalibard, and G. Roger, Phys. Rev. Lett. {\bf 49}, 1804 (1982); G. Weihs, \textit{et al.}, Phys. Rev. Lett. {\bf 81}, 5039 (1998).
\bibitem{102} for instance, M.A. Rowe, \textit{et al.}, Nature (London) {\bf 409}, 791 (2001); D. N. Matsukevich, \textit{et al.}, Phys. Rev. Lett. {\bf 100}, 150404 (2008).
\bibitem{103} D.L. Moehring, \textit{et al.}, Phys. Rev. Lett. {\bf 93}, 090410 (2004); \textit{ibid} {\bf 93}, 109903 (2004).
\bibitem{104} H. Sakai \textit{et al.}, Phys. Rev. Lett. \textbf{97}, 150405 (2006).

\bibitem{12} S. Kochen, and E.P. Specker, J. Math. Mech.  \textbf{17}, 59 (1967).
\bibitem{11} N.D. Mermin, Rev. Mod. Phys.  \textbf{65}, 803 (1993).
\bibitem{10} Y. Hasegawa \textit{et al.}, Nature  \textbf{425}, 45 (2003).
\bibitem{13}D.M. Greenberger, M.A. Horne, and A. Zeilinger, in \textit{Bell's Theorem, Quantum Theory, and Conceptions of the Universe} (ed. M. Kafatos) p.73 (Kluwer Academic, Dordrecht, 1989); D.M. Greenberger \textit{et al.}, Am. J. Phys.  \textbf{58}, 1131 (1990).
\bibitem{27} N.D. Mermin, Phys. Rev. Lett.  \textbf{65}, 1838 (1990).
\bibitem{15} J.W. Pan \textit{et al.}, Nature  \textbf{403}, 515 (2000).
\bibitem{16} Z. Zhao \textit{et al.}, Nature  \textbf{430}, 54 (2004).
\bibitem{105} C.A. Sackett {\it et al.}, Nature (London) {\bf 404}, 256 (2000).
\bibitem{21} A. Rauschenbeutel \textit{et al.}, Science  \textbf{288}, 2024 (2000).

\bibitem{23} Y. Hasegawa \textit{et al.}, Phys. Rev. Lett.  \textbf{97}, 230401 (2006).
\bibitem{24} Y. Hasegawa \textit{et al.}, Phys. Rev. A  \textbf{76}, 052108 (2007).
\bibitem{25} S. Sponar \textit{et al.}, Phys. Rev. A  \textbf{78}, 061604(R) (2008).
\bibitem{26} S. Sponar \textit{et al.}, Phys. Lett. A \textbf{372}, 3153 (2008).
\bibitem{28} H. Rauch, and S.A. Werner, \textit{Neutron Interferometry} (Clarendon, Oxford, 2000).
\bibitem{29} J. Summhammer, Phys. Rev. A  \textbf{47}, 556 (1993).
\bibitem{30} R. Golub, R. Gahler, and T. Keller, Am. J. Phys.  \textbf{62}, 779 (1994).
\bibitem{200} A. Cabello \textit{et al.}, Phys. Rev. Lett. \textbf{100}, 130404 (2008); H. Bartosik \textit{et al.}, Phys. Rev. Lett. \textbf{103}, 040403 (2009).
\bibitem{31} W. D\"{u}r, G. Vidal, and J.I. Cirac, Phys. Rev. A  \textbf{62}, 062314 (2000).
\bibitem{32} S. Filipp \textit{et al.}, Phys. Rev. A \textbf{72}, 021602(R) (2005).

\end{thebibliography}
\end{document}